\documentclass[wssquare]{ws-procs961x669} 

\usepackage[printonlyused]{acronym}
\usepackage{siunitx}

\acrodef{GW}[GW]{gravitational wave}
\acrodef{BBH}[BBH]{binary black hole}
\acrodef{IMBH}[IMBH]{intermediate mass black hole}
\acrodef{IMBHB}[IMBHB]{intermediate mass black hole binary}
\acrodef{IFAR}[IFAR]{inverse false alarm rate}
\acrodef{cWB}[cWB]{coherent waveburst}
\acrodef{NR}[NR]{numerical relativity}
\acrodef{O3}[O3]{third observing run of Advanced LIGO and Advanced Virgo}
\acrodef{LVK}[LVK]{LIGO Scientific,  Virgo \& Kagra collaboration}
\acrodef{SNR}[SNR]{signal-to-noise ratio}
\acrodef{PSD}[PSD]{power spectral density}
\acrodef{CBC}[CBC]{compact binary coalescence}

\begin{document}
\title{Salient features of the optimised PyCBC IMBH search}

\author{Koustav Chandra,  Archana Pai}
\address{Department of Physics, Indian Institute of Technology Bombay, Powai, Mumbai 400 076, India}
\author{V.~Villa-Ortega, T.~Dent}
\address{IGFAE, Campus Sur, Universidade de Santiago de Compostela, 15782 Spain}
\author{C.~McIsaac, I.~W.~Harry, G.~S.~Cabourn~Davies}
\address{University of Portsmouth, Portsmouth, PO1 3FX, United Kingdom}
\author{K.~Soni}
\address{Inter-University Centre for Astronomy and Astrophysics, Pune 411007, India}

\begin{abstract}
Matched-filter based PyCBC searches have successfully detected $\sim \mathcal{O}(50)$ compact binary merger signals in the LIGO-Virgo data. While most PyCBC searches have been designed to detect stellar-mass binaries, we present here a new search that is optimised to detect short-duration \ac{GW} signals emitted by intermediate-mass black hole mergers. 
When applied to the first half of the third observation run data, the optimised search re-identified the \ac{IMBH} binary event, GW190521, with a false alarm rate of 1 in 727 yrs, significantly lower than the previous PyCBC search result of 1 in 0.94 yr. 
Analysis of simulated signals from \ac{IMBH} binaries with generically spinning components shows an increase in sensitivity by a factor of 1.2 to 3 over previous PyCBC searches.
\end{abstract}


\bodymatter

\section{Introduction}\label{sec:intro}
Coalescing compact binaries are amongst the primary sources of \acf{GW} with frequencies accessible to current second-generation ground-based \ac{GW} interferometers. Detection of $\sim \mathcal{O}(50)$ stellar-mass binary mergers (the majority being the binary black holes) has already ushered us into the age of observational strong-field gravity. They have so far given us unique insights into the black hole population that we hardly knew existed, with one of them being the first confirmed detection of a massive black hole system GW190521 with a remnant \ac{IMBH} ~\citep{LIGOScientific:2020iuh, LIGOScientific:2020ufj}.

GW190521 was observed in the first half of the third observing run of the Advanced LIGO and Advanced Virgo detectors.  The compact binary merger event lasted for $\sim 0.1$ \si{s} in the detector band and had barely any observable pre-merger phase of the signal, clearly reflecting the binary's high total mass. The detailed Bayesian parameter estimation showed the merger event to be consistent with the merger of two black holes in a mildly precessing orbit, with component masses of $85^{+21}_{-14}\,M_{\odot}$ and $66^{+17}_{-18}\,M_{\odot}$ and a remnant black hole of $142^{+28}_{-16}\,M_{\odot}$ falling in the mass range of intermediate-mass black holes \cite{LIGOScientific:2020iuh}. The merger event producing this remnant was detected by a search sensitive to generic transient at high confidence with a false alarm rate of $< 1/4900$ years~\citep{Szczepanczyk:2020osv}. 

Traditionally,  the unmodeled GW transient search for short-duration GW signals is performed using the \ac{cWB} search~\citep{Klimenko:2015ypf} which is built on the rationale that a \ac{GW} signal from a \ac{CBC}, chirp or not, is expected to be a transient waveform and it will produce some localised excess of energy if projected in the time-frequency plane. Therefore, identifying such an excess coherently across two or more detectors should give a strong indication of a GW event. Together with this,  model-based matched-filter searches, such as PyCBC~\citep{Allen:2005fk, Usman:2015kfa,DalCanton:2017ala, Nitz:2017svb, Davies:2020tsx} and GstLAL~\citep{Messick:2016aqy}, targeting generic binaries have also been used for the IMBH binary searches in the LIGO-Virgo data.

While unmodeled cWB searches are tuned to look for short-duration signals, \cite{LIGOScientific:2019ysc},  the standard matched-filter-based searches were not.  The lack of \textit{optimisation} towards \ac{IMBH} binary signals made them suffer from the ``look-elsewhere'' effect.  Further, improvements in the sensitivity of the detectors at low frequencies~\citep{LIGO:2021ppb, Fisher:2020, LIGOScientific:2019hgc} before the beginning of the \ac{O3} motivated us to adapt our matched-filter algorithms to the target search space.  Here, we will summarise the salient features of one of the optimised templated searches, namely PyCBC-IMBH that was developed and deployed for the \ac{IMBH} binary search. This optimised search was used together with other search pipelines to look for \ac{IMBH} binaries in the \ac{O3} run, and the reader can find the results of it in \citep{LIGOScientific:2021tfm}. For complete information on the algorithm itself, we direct the reader to the optimised PyCBC \ac{IMBH} search paper \cite{Chandra:2021wbw}.

\section{The optimised PyCBC-IMBH search}\label{sec:optimise}

Below we summarise the salient features of the optimised PyCBC-IMBH search \cite{Chandra:2021wbw}. 

{\it Data cleaning:} The gravitational wave strain data is often plagued with loud background noise, which can significantly hamper the search sensitivity~\citep{Cabero:2019orq}.  Thus, we begin our analysis by windowing out very high amplitude excursions ($ > 50 \sigma$ deviation from Gaussian noise) in the whitened data~\citep{Usman:2015kfa}. Since empirically, these loud noise transients tend to be correlated with neighbouring quieter noise transients, we also \textit{discard} any LIGO triggers which are either 1s before or 2.5s after the centre of the gating window~\citep{Chandra:2021wbw}.

{\it Template bank}: After this data cleaning process, we use matched-filtering to detect the target signals hidden in the detector noise~\citep{Sathyaprakash:1991mt, Owen:1995tm, Allen:2005fk}. We assume that our target signals are well-modelled and deterministic, and to a good approximation, we can consider them to be buried in wide-sense stationary Gaussian noise. However, the binary's parameters are not known a priori, forcing us to use a discrete bank of filter waveforms, aka templates. Our bank of templates is designed to intercept the dominant quadrupolar harmonics of a non-precessing binary~\citep{Ajith:2009bn}. Specifically, we use the reduced-order representation of the spinning effective one-body model, \texttt{SEOBNRv4}~\cite{Bohe:2013cla, Purrer:2015tud}, for our templates, and we \textit{constrain} ourselves to binaries with red-shifted (detector frame) total mass between $100-600 M_\odot$~\citep{Chandra:2021wbw}. Additionally, our templates have a minimum component mass of $40 M_\odot$ and a mass ratio between 1/1 and 1/10. We do not use relatively lower mass ratios in the templates as binaries with low mass ratios emit gravitational wave signals with significantly higher sub-dominant harmonics that are not modelled by our waveform approximant. We also don't include any templates with a duration less than 0.07 s., measured from a fixed starting frequency of 15 Hz, to reduce the false alarms arising from short-duration noisy transients. Since information on the spins of \ac{IMBH} binaries is lacking, we allow our templates to have dimensionless spin components along the orbital angular momentum to be between $\pm 0.998$. The reader can find additional details about the template bank in \citep{Chandra:2021wbw}.

Although our search targets non-precessing binaries, we expect our search to retain its sensitivity towards moderately precessing binaries, mainly because our target systems are relatively short-lived within the detector bandwidth~\citep{Harry:2016ijz, Chandra:2020ccy}. Furthermore, we also expect our search to be equally sensitive to binaries with low eccentricity as the emission of gravitational waves will circularise the binary system before the merger, and because of the shortness of our templates and the detector bandwidth, we will only be able to intercept the last few cycles of \ac{IMBH} binary's inspiral~\citep{Ramos-Buades:2020eju}.

{\it Time-phase consistency between detectors}: Single-detector triggers are identified based on peaks in the matched filter \ac{SNR} time series. Unfortunately, our detector data is polluted with several non-stationary instrumental artefacts that are not loud enough to be gated out during data pre-processing. Technically, this makes our search a detection problem in non-Gaussian noise. Having a network of simultaneously operating similar sensitivity detectors help a lot as we can demand that the time difference between triggers at different detectors is consistent with an astrophysical signal, i.e. within the light-travel time up to expected uncertainty.

Besides, if the trigger is due to a \ac{GW} signal, then it must share consistent phase and amplitude characteristics. Hence the \textit{coincidence test} significantly reduces the number of noisy artefacts that is picked up by the search. 

{\it Signal consistency tests:} But there can be chance coincidences between noise triggers. So, before the coincidence test is performed, there is a need to differentiate a signal from a glitch at the single detector level. We do this in multiple stages. We begin by only considering the loudest trigger within a pre-determined window of time to reduce false alarms, and we subject these triggers to two different signal-consistency tests: the $\chi_r^2$-test~\citep{Allen:2004gu} and the $\chi_{r,sg}^2$-test~\citep{Nitz:2017lco}. The former checks if the morphology of the triggers matches with that of the best-matched template, while the latter checks for any excess power beyond the maximum frequency of the template. The output of these two tests is used to amend the trigger \ac{SNR} to reduce the apparent loudness of a noise trigger as the target signal will return values closer to unity. We also account for any short-term non-stationarity of the detector noise that can lead to wrongful estimation of the trigger \ac{SNR}~\citep{Venumadhav:2019lyq, Venumadhav:2019tad, Mozzon:2020gwa}.

Given our target space, our search is susceptible to short but loud noise triggers motivating us to use stricter discriminators. We chose to discard any trigger with $\chi_r^2 > 10$ or with short-term \ac{PSD} measure of over ten times the expectation from stationary noise~\citep{Chandra:2021wbw}. 

{\it Event significance and ranking:} We subject the surviving single-detector triggers to the coincidence test and assign each\textit{ multi-detector candidate} a rank based on their response to the coincidence test~\citep{Nitz:2017svb, Davies:2020tsx}. Our candidate events can be coincidentally observed either in one of the possible two detector combinations or in all three detectors. This is because we analyse all possible two and three detector combinations. Finally, a statistical significance is assigned to each of these candidates by comparing its rank against simulated background noise that is generated by finding fictitious coincidences between detector data~\citep{Usman:2015kfa}. We achieve this by shifting the pivot detector(s) data with respect to a fixed reference detector by a time that is considerably larger than the light-travel time between the detectors. By repeating this procedure, we produce more than $10^{4}$ years worth of noise background.

In the next section, we provide details of how the PyCBC algorithm benefits from the optimisation made.
\begin{table}
\tbl{Summary of the single pipeline sensitive volume-time $\langle VT \rangle$. Here, $M_T$ is the total source frame mass of the binary and $q=m_2/m_1 \leq 1$ denotes the mass ratio of the binary. $\chi_\mathrm{eff}$ is the mass-weighted combination of the spins parallel to the orbital angular momentum $\vec{L}$ while $\chi_\mathrm{p}$ captures the average amount of precession exhibited by a generically precessing system over many cycles defined at reference frequency of 11\,Hz during the inspiral.}
{\begin{tabular}{@{}ccccccc@{}}
\toprule
$M_T~(M_\odot)$ & $q$ & $\chi_\mathrm{eff}$ & $\chi_\mathrm{p}$ & $\langle VT \rangle_{\text{PyCBC}}$[\si{Gpc^3yr}] & $\langle VT \rangle_{\text{cWB}}$[\si{Gpc^3yr}] & $\langle VT \rangle_{\text{GstLAL}}$ [\si{Gpc^3yr}] \\
\colrule
120 &  1/2 &    0.00 &  0.00&  11.51 &   8.93 &   8.18 \\
   120 &  1/4 &    0.00 &  0.00&  4.64 & 3.54 &   3.04 \\
   120 &  1/5 &    0.00 &  0.00&  3.11 & 2.42 &   2.06 \\
   120 &  1/7 &    0.00 &  0.00&  1.67 & 1.24 &   1.05 \\
   120 & 1/10 &    0.00 &  0.00&  0.82 & 0.60 &   0.51 \\
   150 &  1/2 &    0.00 &  0.00 & 11.95& 9.96 &   7.40 \\
   200 &    1 &    0.00 &  0.00&  14.84 & 12.76 & 10.27 \\
   200 &  1/2 &    0.00 &  0.00&  10.35 &   9.47 & 6.56 \\
   200 &  1/4 &    0.00 &  0.00&  3.90 & 3.69 & 1.93 \\
   200 &  1/7 &    0.00 &  0.00&  1.34 & 1.29 & 0.53 \\
   220 & 1/10 &    0.00 &  0.00&  0.57 & 0.60 & 0.22 \\
   250 &  1/4 &    0.00 &  0.00&  2.99 & 2.83 & 1.80 \\
   300 &  1/2 &    0.00 &  0.00&  6.27 & 5.86 & 5.38 \\
   350 &  1/6 &    0.00 &  0.00&  0.72 & 0.75 & 0.44 \\
   400 &    1 &    0.00 &  0.00&  4.82 & 4.55 & 4.29 \\
   400 &  1/2 &    0.00 &  0.00&  3.08 & 3.33 & 2.68 \\
   400 &  1/3 &    0.00 &  0.00&  1.68 & 1.94 & 1.29 \\
   400 &  1/4 &    0.00 &  0.00 & 1.11 & 1.22 & 0.70 \\
   400 &  1/7 &    0.00 &  0.00 & 0.44 & 0.47 & 0.20 \\
   440 & 1/10 &    0.00 &  0.00 & 0.20 & 0.20 & 0.09 \\
   500 &  2/3 &    0.00 &  0.00 & 1.87 & 1.97 & 1.45 \\
   600 &    1 &    0.00 &  0.00&  0.90 & 0.79 & 0.50 \\
   600 &  1/2 &    0.00 &  0.00&  0.65 & 0.81 & 0.34 \\
   800 &    1 &    0.00 &  0.00&  0.15 & 0.14 &  0.11 \\
   200 &    1 &    0.80 &  0.00&  38.54 & 29.60 & 30.99 \\
   400 &    1 &    0.80 &  0.00&  18.08 & 15.88 & 16.07 \\
   600 &    1 &    0.80 &  0.00&  4.97 & 5.35 &  4.03 \\
   800 &    1 &    0.80 &  0.00&  0.89 & 0.88 &  0.37 \\
   200 &    1 &   -0.80 &  0.00&  10.07 & 9.28 & 6.39 \\
   400 &    1 &   -0.80 &  0.00&  1.90 &  1.86 & 1.66 \\
   600 &    1 &   -0.80 &  0.00&  0.22 &  0.22 & 0.11 \\
   800 &    1 &   -0.80 &  0.00&  0.04 &  0.05 &  0.03 \\
   200 &    1 &    0.51 &  0.42&  26.38 & 20.88 & 19.42 \\
   200 &  1/2 &    0.14 &  0.42&  13.97 & 12.36 &  9.01 \\
   200 &  1/4 &    0.26 &  0.42&  7.67 & 7.65 &  4.13 \\
   200 &  1/7 &    0.32 &  0.42&  3.35 & 3.59 &  1.44 \\
   400 &    1 &    0.51 &  0.42&  10.53 & 9.38 & 9.19 \\
   400 &  1/2 &    0.14 &  0.42&  4.96 & 5.42 &  4.24 \\
   400 &  1/4 &    0.26 &  0.42&  2.58 & 3.63 &  1.88 \\
   400 &  1/7 &    0.32 &  0.42&  1.11 & 1.79 &  0.65 \\
   600 &    1 &    0.51 &  0.42&  2.44 & 2.49 &  1.72 \\
   600 &  1/2 &    0.14 &  0.42&  1.12 & 1.45 & 0.58 \\
   800 &    1 &    0.51 &  0.42&  0.10 & 0.18 &  0.01 \\
\botrule
\end{tabular}
}
\label{tab:main}
\end{table}

\section{Summary of Sensitivity Comparison}\label{sec:results}

To quantify the benefit of our restricted analysis, we compared the sensitivity of our optimised search against existing PyCBC-based searches with overlapping parameter space~\cite{Chandra:2021wbw}. We compared the PyCBC-IMBH search against the PyCBC-broad and the PyCBC-BBH search. The former looks for binaries consisting of either neutron stars or black holes or both, while the latter is a ``focused'' search for \ac{BBH} covering a restricted range of masses. We resort to an injection campaign for the comparison. This involved adding simulated signals to real \ac{GW} data from the first half of the third observing run (O3a) run as part of the analysis without any physical actuation and then estimate the sensitivity of each of the searches in terms of sensitive volume time, $\langle VT \rangle$~~\citep{Biswas:2007ni, LIGOScientific:2016hpm}. Our simulated signals mimicked those from generically spinning \ac{BBH} with detector frame total mass between $100-600 M_\odot$ and having a mass ratio between $1/1-1/10$. The signals themselves were either the dominant quadrupolar mode or the full symphony emitted by such binaries. We found that the PyCBC-IMBH search is $\sim 1.2-3$ times more sensitive than the PyCBC-broad search and is $\sim 1.1 - 12.6$ times more sensitive than the PyCBC-BBH search at a false alarm rate threshold of 1 in 100 years.

We also compare our search against two other searches, namely \ac{cWB} and GstLAL-IMBH, that were used by the \ac{LVK} to look for \ac{IMBH} binaries in \ac{O3} run~\cite{LIGOScientific:2021tfm}. For this purpose, we utilise the results of the injection campaign that were used by the \ac{LVK} to place upper limits on the merger rate density for a discrete set of \ac{IMBH} binaries. The dataset is publicly available at \url{https://dcc.ligo.org/LIGO-P2100179/public}, and it contains details about the parameters of each of the injections and also their associated p-value as reported by the contributing searches. The signals themselves were numerically simulated and are designed to imitate \acp{GW} from a representative population of \ac{IMBH} binaries with generic spins. Additional details about the injections can be found in \citep{LIGOScientific:2021tfm}. 

To estimate $\langle VT \rangle$ using the publicly available dataset, we start off by calculating the number of signals that are recovered by each of three contributing searches. We do this by considering only those injections whose p-value is less than the minimum p-value of the loudest noise trigger, namely that of 200214\_224526 in the \citep{LIGOScientific:2021tfm}. This trigger has a minimum p-value of $9.2\times 10^{-2}$ and was only reported by the weakly model-dependent \ac{cWB} search. The $\langle VT \rangle$ is then given by
\begin{equation}
    \langle VT \rangle = \frac{N_\text{rec}}{N_\text{total}}\langle VT \rangle_\text{total}
\end{equation}
where the first term denotes the fraction of injections that are recovered below the p-value threshold while the second term denotes the total spacetime volume covered by a given injection set.

We tabulate the results of this analysis for each of the three pipelines for each simulated \ac{IMBH} binary in Table~\ref{tab:main}. Amplitude and phase errors arising from detector calibration have not been included in the analysis, but a statistical uncertainty of 4\%-7\% is expected. We find that the PyCBC-IMBH search has higher sensitivity as compared to the GstLAL-IMBH for all of the mass bins and is comparably sensitive (if not better for most of the symmetric and/or relatively low total mass bins) to \ac{cWB} search.

The latter can be understood as follows. As discussed in Sec.~\ref{sec:optimise}, the PyCBC-IMBH search uses only the dominant multipole templates in matched filtering. Hence, the sensitivity drops with a decrease in mass ratio as compared to the \ac{cWB} search. This is because a decrement in mass ratio corresponds to an increase in the contribution coming from the sub-dominant harmonics. The \ac{cWB} search being a weakly modelled search, is oblivious to this effect and hence has higher sensitivity than template-based searches. Furthermore, if one compares the sensitivity of a given search for a given total mass but decreasing mass ratio, one will observe that the search performance degrades owing to an overall decrease in the intrinsic luminosity of the system involved.

Concerning the total mass of a binary, the PyCBC search tends to have higher sensitivity for relatively lower total mass as signals from such systems tend to have a longer inspiral within the detector bandwidth, thus allowing for a larger \ac{SNR}. At higher total mass, the \ac{cWB} search has higher sensitivity because of its efficient signal-noise discriminators, which allow it to distinguish the signal from noisy transients better, thus allowing for a relatively lower p-value and hence greater sensitivity. This also explains why the PyCBC search has a poorer performance for a system with a given mass distribution but having a relatively anti-aligned spin.

\ac{GW} signals from a significantly precessing binary can have a notably suppressed inspiral right before the merger, causing a characteristic sine-Gaussian morphology. The search summarised here uses non-precessing quadrupolar templates meaning the \ac{SNR} recovered would be lower and the $\chi_{r}^2$ higher owing to mismatch with templates.

\section{Conclusion}
Here, we summarise the salient features of the optimisation made to the existing PyCBC search to target short-duration \acp{GW} from massive \acp{BBH}.  This optimisation enabled us to improve the search performance and re-detect the event GW190521 with improved statistical significance. We also compare the sensitivity of PyCBC-IMBH with other searches used for the detection of \ac{IMBH} binaries using the public data. 

Detection of \ac{IMBH} and its subsequent studies has the potential to provide valuable insight into the \ac{BBH} population. Hence searches like this are going to be increasingly important. While we are primarily concerned here with detection of \ac{IMBH} binaries, this type of search can be used to detect highly red-shifted \acp{BBH} which we expect to see in the next few years with the advent of third-generation \ac{GW} observatories like the Einstein Telescope and Cosmic Explorer. Thus, the future of this field is quite exciting! 

\section{Acknowledgements}
KC is extremely grateful to the organisers of the conference for hosting such a wonderful meeting in challenging circumstances. We thank Gayathri V. and Keith Riles for their comments on an earlier draft of this proceeding. The authors are grateful for the computational resources and data provided by the LIGO Laboratory which is funded by National Science Foundation Grants No. PHY-0757058 and No. PHY-0823459. The open data is available in the Gravitational Wave Open Science Center (https://www.gw-openscience.org/), a service of LIGO Laboratory, the LIGO Scientific Collaboration and the Virgo Collaboration. The authors also acknowledge the use of the IUCAA LDG cluster, Sarathi, for computational/numerical work. KC acknowledges the MHRD, Government of India, for the fellowship support.  AP's research is supported by SERB-Power-fellowship grant SPF/2021/000036, DST, India. 
VVO and TD acknowledge financial support from Xunta de Galicia (Centro singular de investigación de Galicia accreditation 2019-2022), by European Union ERDF, and by the ``María de Maeztu'' Units of Excellence program MDM-2016-0692 and the Spanish Research State Agency.  GSCD and IWH acknowledge the STFC for funding through grant ST/T000333/1. CM was supported by the STFC through the DISCnet Centre for Doctoral Training.  KS acknowledges the Inter-University Centre of Astronomy and Astrophysics (IUCAA), India, for the fellowship support. This document has LIGO preprint number LIGO-P2100352.

\textit{We want to thank all of the essential workers who put their health at risk during this ongoing COVID-19 pandemic. Without their support, we would not have completed this work. We offer condolences to people who have lost their family members during this pandemic.}

\bibliographystyle{ws-procs961x669}
\bibliography{ws-pro-sample}

\end{document}